\documentclass{iopart}
\usepackage{graphicx}
\usepackage{dcolumn}
\usepackage{bm}
\usepackage{float}
\usepackage{color}

\begin{document}

\title{Thermodynamic properties of titanium from ab-initio calculations}
\author{Uri Argaman$^1$,Eitan Eidelstein$^2$,Ohad Levy$^2$,Guy Makov$^1$}
\address{$^1$Materials Engineering Department, Ben-Gurion University of the Negev, Beer Sheva 8410501, Israel}
\address{$^2$Department of Physics, NRCN, P.O. Box 9001, IL Beer-Sheva, 84190, Israel}
\ead{makovg@bgu.ac.il}

\begin{abstract}
The lattice parameters, lattice stability and phonon dispersion curves of five proposed phases of Ti: $\alpha$, $\beta$, $\gamma$, $\delta$ and $\omega$ are investigated within DFT. It is found that the sequence of high pressure phases at zero temperature is  $\alpha \rightarrow \omega \rightarrow \gamma \rightarrow \delta \rightarrow \beta$ with the $\delta$ and $\beta$ phases becoming degenerate at high pressure. However, the $\gamma$ phase may be unstable as is reflected  by the existence of imaginary values in the phonon spectra. The results of the DFT calculations are employed to estimate the entropy and free energies of the $\alpha$ and $\omega$ phases. It is found that converged phonon calculations lead to an entropy difference which is much smaller than previous estimates, and a much steeper $\alpha-\omega$ phase transition line. 
\end{abstract}


\section{Introduction.}\label{sec.intro}
Titanium has long been known to have three stable phases at low pressures: ~\cite{young} $\alpha$ which is stable at room temperature and pressure, $\omega$ which is stable at higher pressures and $\beta$ which is stable at higher temperatures. Recently, two additional high pressure phases of titanium at room temperature were discovered: $\delta$ and $\gamma$ ~\cite{gamma,delta} and one phase, which is reportedly stable only at high pressure and high temperature, $\eta$ ~\cite{eta}. Despite extensive studies, the phase diagram of titanium in the pressure-temperature plane retains a large uncertainty, in particular with respect to the location of the $\alpha$-$\omega$ phase transition line ~\cite{Errandonea,Zhang,Tonkov}. This uncertainty has attracted extensive theoretical interest claiming to predict the locus of the phase transition line.

Experimentally, the $\alpha$-$\omega$ phase transition at room temperature is observed to occur across a wide pressure range between 2.9 GPa to 11 GPa ~\cite{Errandonea}, where it is suggested that this large range is due to deviations from hydrostatic conditions, hysteresis phenomena, phase coexistence and it depends on the pressure medium employed. The temperature dependence of the phase transition pressure also retains a large uncertainty. From the Clausius-Clapeyron relation, the slope of the phase transition pressure in the P-T plane is determined by the ratio of the entropy difference and volume difference between the two phases. The entropy difference between the $\alpha$ and $\omega$ phases is not known accurately. Zhang et al. ~\cite{Zhang} estimate the entropy difference to be 0.57 $J/mol/K$ as opposed to Tonkov and Ponyatovsky ~\cite{Tonkov} who reported a difference of 1.49 $J/mol/K$. For the volume change between $\alpha$ and $\omega$ phases, the agreement between different experiments is better: Zhang et al. ~\cite{Zhang} determine the average volume change between $\alpha$ and $\omega$ phases to be 0.197 $cm^3/mole$ and Tonkov and Ponyatovsky ~\cite{Tonkov} reported this volume change to be 0.138 $cm^3/mole$.

Titanium and titanium group elements have also been the subject of several first principle studies, including calculations of the phonon spectrum of titanium as well as of thermodynamic properties based on these calculations. Total energy density functional theory (DFT) calculations of Ti, Zr and Hf at zero Kelvin by Ahuja et al. ~\cite{covalent} found that the $\omega$ phase of titanium is stable at zero temperature and zero pressure. These results have been reproduced in subsequent investigations. They suggest that in the $\omega$ phase there is a substantial covalent contribution to the bonding. Nie et al. ~\cite{Nie} calculated thermodynamic properties of titanium metal and found that the contribution of electronic excitations to the heat capacity of titanium is not negligible. In addition, they calculated the phononic contribution to the heat capacity, linear thermal expansion and bulk modulus using density functional perturbation theory (DFPT) ~\cite{DFPT}  without presenting results for the phonon spectrum.

Recently, Mei et al. ~\cite{Liu} reported  free energy calculations of thermodynamic properties for the $\alpha$,$\beta$ and $\omega$ phases of titanium. Phonon results were combined with total energy at zero Kelvin and electronic contributions to calculate the entropy and free energies. The entropy difference between the $\alpha$ and $\omega$ phases was found to be more than 2 $J/mol/K$ at zero pressure and the volume difference to be 0.138 $cm^3/mole$. The phase transition between the $\alpha$ and $\omega$ phases at zero pressure was estimated to be at 186 Kelvin. Hu et al. ~\cite{Alfe} also investigated the thermodynamic properties of titanium and found good agreement with experimental data of heat capacity, linear thermal-expansion and bulk modulus in the $\alpha$ phase across a wide range of temperatures. They calculated the phase-diagram of titanium using both the quasiharmonic Debye model and directly from the phonon DOS and found that there is no significant difference between these two methods. The entropy difference between the $\alpha$ and $\omega$ phases was found to be 1.3 $J/mol/K$ at zero pressure and the volume difference was found to be 0.144 $cm^3/mole$ with the phase transition occurring at 146 K. Within the quasiharmonic Debye model they also calculated the thermodynamic properties of the $\beta$ phase, which is mechanically unstable at zero Kelvin. In both these studies the authors calculated the phonon spectrum using the small displacement method  ~\cite{sdm} but did not report on their convergence with respect to the supercell size or number of q-points.

Despite the wide range of experimental and theoretical works present in the literature, there remains a considerable uncertainty in the $\alpha$-$\omega$ phase transition pressure at zero Kelvin as well as in the temperature dependence of the phase transition pressure. From extrapolation of the phase transition line in the work of Zhang et al. ~\cite{Zhang} it is seen that the $\alpha$ phase is stable at zero pressure and zero Kelvin. This result is in contradiction to the zero temperature DFT calculations, raising a possible question of accuracy in determining the stable phase. In addition, the entropy difference between the two phases presents a large uncertainty. Because experimentally determined phase transition pressures are not accurately known it is difficult to compare between experiment and first principle calculations and estimate the error of such calculations. In particular, it is difficult to separate between the error in the total energy, arising from the exchange-correlation and pseudopotential approximations, and between that of the phonon calculations, that add an additional source of error due to the quasi-harmonic approximation and first order perturbation theory.

At high pressure, there is no agreement in the literature on the stable phases of titanium reported in several experimental investigations of this subject. Vohra and Spencer ~\cite{gamma} observe a transformation from the $\omega$ phase to an orthorhombic phase $\gamma$ at $116\pm4$ GPa. Akahama et al. ~\cite{delta} observe this transformation at $128-130$ GPa. As pressure is increased, they found an additional transformation from the $\gamma$ phase to another orthorhombic phase with the same space group: $Cmcm$, denoted $\delta$. The authors mention that it is not possible to distinguish from the diffraction pattern between the space groups $Cmcm$ and $Cmc2_1$ and they chose the former because it has a higher symmetry. The $\delta$ phase remains stable up to $216$ GPa which is the maximum pressure in their experiment. The $\beta$ phase was not observed in this work at high pressure. So, the sequence of the phases of titanium under pressure at room temperature according to Akahama et al. is $\alpha \rightarrow \omega \rightarrow \gamma \rightarrow \delta$. Ahuja et al. ~\cite{eta} observe a transformation from the $\omega$ phase to the $\beta$ phase (bcc) at a pressure of 40-80 GPa without any intermediate phases, leading to the sequence $\alpha \rightarrow \omega \rightarrow \beta$. Laser-heating at 1200-1300 K or electrical heating at 1050 K at 78-80 GPa results in a transformation from the $\beta$ phase to an orthorhombic phase with $Fmmm$ space group denoted $\eta$ ~\cite{eta}.

The high pressure phases of titanium have also been studied theoretically with DFT calculations. The transformation from the $\omega$ phase to the $\gamma$ phase has been calculated to occur at a pressure between 98-106.3 GPa ~\cite{Hao,Kutepov,Verma} and the transformation from the $\gamma$ to $\delta$ phases is calculated to occur between 106-134.9 GPa ~\cite{Hao,Kutepov}. At higher pressures, some theoretical works predict a transformation from the $\delta$ phase to the $\beta$ phase at 136-160.8 ~\cite{Hao,Kutepov}. In contrast to these results, Mei et al. ~\cite{Liu2}. predict that the $\delta$ phase is not stable under hydrostatic compression at all and the sequence of the phases of titanium under pressure is $\alpha \rightarrow \omega \rightarrow \gamma \rightarrow \beta$. 

In the present contribution we present fully relaxed zero Kelvin DFT calculations for the phases $\alpha$, $\omega$, $\beta$, $\gamma$ and $\delta$, and phonon calculations obtained using DFPT. The sequence of high pressure phases at zero Kelvin is determined to be $\alpha \rightarrow \omega \rightarrow \gamma \rightarrow \delta \rightarrow \beta$. From the phonon calculations we find that the proposed $\gamma$ phase is mechanically unstable at zero Kelvin. These calculations are employed to determine the thermodynamic properties of pure titanium in the stable phases. The convergence of the thermodynamic calculations with respect to the number of q-points in the phonon calculations is examined and we find that previous calculations were probably insufficiently converged. The achievable accuracy in calculations of phase transition is discussed, with $\alpha$ and $\omega$ phases of titanium as an example. 

\section{Theory.}\label{sec.theory}
Within the Born-Oppenheimer approximation, the Helmholtz free energy can be separated into three terms:

\begin{equation}
F(V,T)=E(V)+F_{vib}(V,T)+F_{el}(V,T)
\label{F}
\end{equation}
where $E(V)$ is the zero temperature energy, $F_{vib}(V,T)$ is the vibrational contribution to the total energy and $F_{el}(V,T)$ is the contribution of electronic excitations to the free energy. 

Within the quasiharmonic approximation (QHA), $F_{vib}(V,T)$ is the sum of free energies of quantum harmonic oscillators at constant volume and temperature:  

\begin{eqnarray}
F_{vib}(V,T)= \nonumber\\
k_{B}T\int_{-\infty}^{\infty}ln\left(2\cdot sinh\left(\frac{\hbar\omega}{2k_{B}T}\right)\right)\cdot g(\omega,V)d\omega
\end{eqnarray}
where $g(\omega,V)$ is the phonon DOS and $\omega$ is the phonon frequency. $F_{el}(V,T)$ can be determined from the electronic DOS:

\begin{equation}
F_{el}(V,T)=E_{el}(V,T)-T\cdot S_{el}(V,T)
\end{equation}
where $E_{el}(V,T)$ is the internal energy of electronic excitations defined by:

\begin{eqnarray}
E_{el}(V,T)=  \nonumber\\
\int n(E,V)\cdot E\cdot f(E)\cdot dE-\int_{-\infty}^{E_{f}}n(E,V)\cdot E\cdot dE
\end{eqnarray}
where $n(E,V)$ is the electronic DOS, $E_{f}$ is the Fermi energy and $f$ is the Fermi-Dirac distribution. $S_{el}(V,T)$ is the entropy of electronic excitations:
\begin{eqnarray}
S_{el}(V,T)= \nonumber\\
-k_{B}\int n(E,V)[fln(f)-(1-f)ln(1-f)]dE
\end{eqnarray}
where $k_{B}$ is the Boltzmann constant. In the present work the electron density of states is approximated by the Kohn-Sham density of states.

Thermodynamic properties can also be calculated using the Debye model, which assumes a simplified, parabolic form of the phonon density of states as a function of the frequency leaving only one parameter to specify the material, known as the Debye temperature. The Debye temperature can be calculated using thermodynamic properties such as the heat capacity or from the elastic constants. One can implement both methods using ab-initio calculations or experimental data. To determine the Debye temperature, $\theta _D$, from the phonon heat capacity as a function of temperature we use a least squares fit. To determine the Debye temperature from the elastic constants we use the relations~\cite{Shukla}:
\begin{equation}
\theta_{D}=\frac{h}{k}\left(\frac{9}{4\pi}\right)^{1/3}V_{atom}^{1/6}\cdot\left(\frac{B}{M_{atom}}\right)^{1/2}\cdot f(\sigma)
\end{equation}
where
\begin{equation}
f(\sigma)=\left(\left(\frac{1+\sigma}{3(1-\sigma)}\right)^{3/2}+2\cdot\left(\frac{2(1+\sigma)}{3(1-2\sigma)}\right)^{3/2}\right)^{-1/3}
\end{equation}
h is the Plank constant, k is the Boltzmann constant, $V_{atom}$ is the volume per atom, B is the bulk modulus, $M_{atom}$ is the mass per atom and $\sigma$ is Poisson's ratio.

The electronic calculations were performed using an ultrasoft pseudopotential plane-waves method ~\cite{Martin}. Phonon spectra were calculated using DFPT ~\cite{DFPT}. All calculations were performed with the Quantum Espresso package ~\cite{QE}. The exchange-correlation is approximated by Becke-Perdew general gradient approximation (GGA) ~\cite{Perdew,Becke} and/or the PBE functional ~\cite{PBE}.  The kinetic energy cutoff for the wave function is 50 $Ry$, the kinetic energy cutoff for charge density and potential is 500 $Ry$, the number of k-points is 16 in each direction of the reciprocal lattice vectors (Monkhorst-Pack algorithm) centered at the $\gamma$ point except in the $\beta$ phase where 24 k-points in each direction of the reciprocal lattice vectors were used. The pseudopotential employed is Ti.bp-sp-van\_ak.UPF from http://www.quantum-espresso.org with 12 valence electrons included. Methfessel-Paxton smearing scheme with a Gaussian spreading of 0.01 $Ry$ was used for the occupation. All these numerical parameters are required for convergence of the total energy to $10^{-5}$ $Ry/atom$. All the geometrical parameters of the lattice are determine with fully relaxation, both for the unit cell shape and the atomic positions. Phonon calculations have been performed with 4 q-points in each direction of the reciprocal lattice vectors in the calculations of $\alpha$, $\omega$ and $\beta$ at zero pressure, 3 q-points for the gamma and delta phases and 8 q-points for the beta phase.
\section{Results}\label{sec.results}
\subsection{Structural properties}

\begin{table}
\caption{Lattice parameters ($\AA$ngstrom) of $\alpha$ and $\omega$ at zero pressure calculated with several exchange-correlation approximations ($a_c$). Experimental data ($a_e$) from reference ~\cite{Zhang2} for $\alpha$ and $\omega$ and from reference ~\cite{Ogi} for $\beta$.}
\begin{indented}
\item[]\begin{tabular}{@{} l l l l l l l }
\br
 & $a_c$ & $a_e$ & $\Delta a$ & $\left(c/a\right)_c$ & $\left(c/a\right)_e$ & $\Delta\left(c/a\right)$\\
\mr
$\alpha$(BP) & 2.93 & 2.95 & -0.7\% & 1.58 & 1.588 & -0.5\\
$\omega$(BP) & 4.56 & 4.61 & -1.1\% & 0.62 & 0.615 & 0.6\\
$\beta$(BP) & 3.24 & 3.31 & -2.1\% & 1 &  & \\
$\alpha$(PBE) & 2.94 & 2.95 & -0.34\% & 1.58 & 1.588 & -0.5\\
$\omega$(PBE) & 4.58 & 4.61 & -0.65\% & 0.62 & 0.615 & 0.6\\
$\alpha$(LDA) & 2.86 & 2.95 & -3.15\% & 1.58 & 1.588 & -0.5\\
$\omega$(LDA) & 4.45 & 4.61 & -3.59\% & 0.62 & 0.615 & 0.8\\
\br
\end{tabular}
\label{Lattice_parameters}
\end{indented}
\end{table}

\begin{table}
\caption{Lattice parameters (Angstrom) of the high pressure phases of titanium. Experimental data from reference ~\cite{delta}.}
\begin{indented}
\item[]\begin{tabular}{@{} l l l l l l l }
\br
 & $a_c$ & $b_c$ & $c_c$ & $a_e$ & $b_e$ & $c_e$\\
\mr
$\gamma$(130 GPa) & 2.351 & 4.458 & 3.855 & 2.382 & 4.461 & 3.876\\
$\delta$(178 GPa) & 3.805 & 2.630 & 3.595 & 3.861 & 2.630 & 3.632\\
$\beta$(230 GPa) & 2.544 &  &  &  &  & \\
\br
\end{tabular}
\label{lattice_parameter_high_pressure}
\end{indented}
\end{table}

\begin{figure}
\centering
        \includegraphics[scale=0.4]{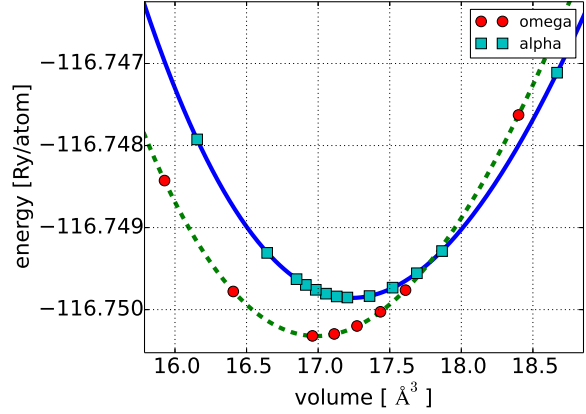}
        \caption{total energy curve of $\alpha$ and $\omega$ phases. Solid lines are third order polynomial fits.}\label{cold_curve}
\end{figure}

\begin{figure}
\centering
        \includegraphics[scale=0.4]{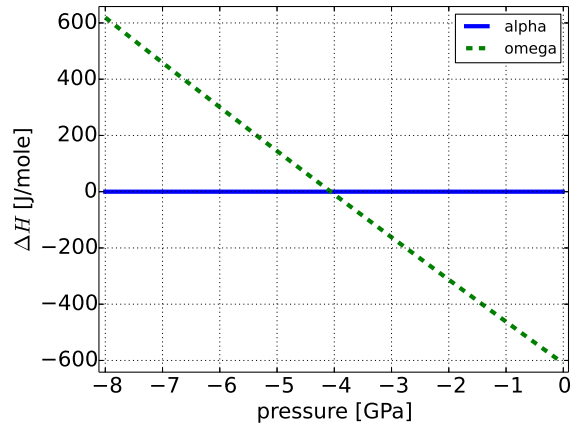}
        \caption{Differences in the enthalpy relative to $\alpha$ phase at zero Kelvin. The phase transition at 0 $K$ without zero-point energy is at -4.06 $GPa$.}\label{enthalpy}
\end{figure}

\begin{figure}
\centering
        \includegraphics[scale=0.4]{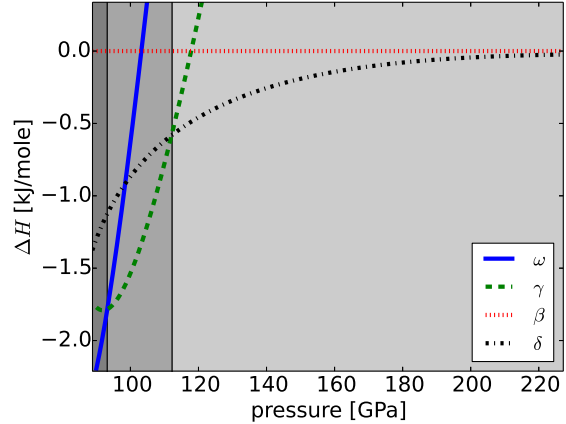}
        \caption{Differences in the enthalpy relative to $\omega$ phase at zero Kelvin. The phase transitions $\omega \rightarrow \gamma \rightarrow \delta \rightarrow \beta$ at 0 $K$ without zero-point energy are at 93 and 112 GPa respectively. At high pressure $\delta$ and $\beta$ become degenerate.}\label{enthalpy_high_pressure}
\end{figure}

\begin{figure}
\centering
        \includegraphics[scale=0.4]{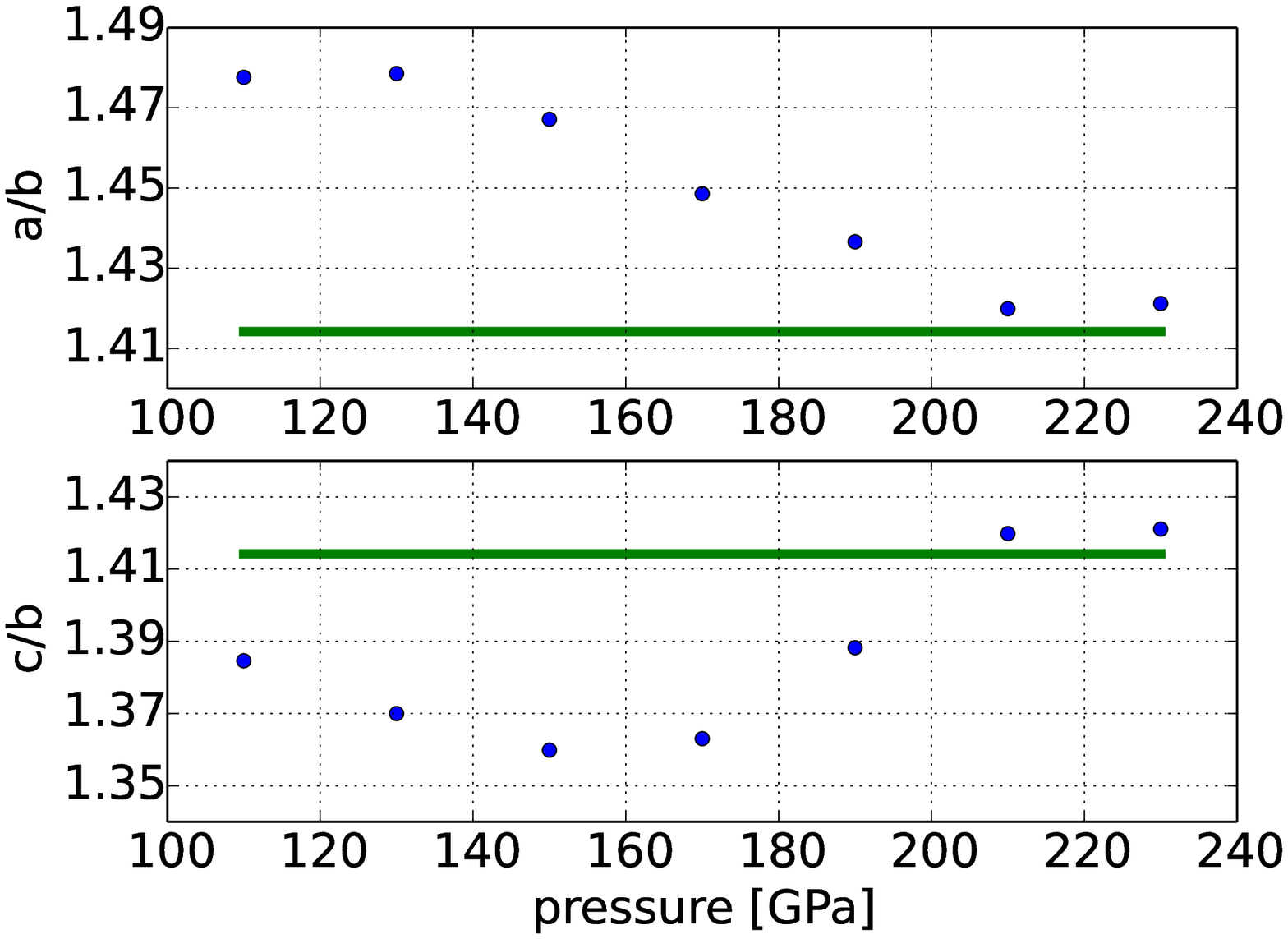}
        \caption{Lattice parameters of $\delta$ phase as a function of the pressure. Solid lines are the bcc values of $\sqrt{2}$.}\label{lattice_par}
\end{figure}

The lattice parameters of the $\alpha$ and $\omega$ phases at zero pressure were determined by relaxation of the unit cell to hydrostatic conditions. From table ~\ref{Lattice_parameters} we can see that there is good agreement with experimental data for the lattice parameters in both $\alpha$ and $\omega$ phases. For the $\beta$ phase there is a slightly larger error relative to experimental data, which might be explained by the high temperature of the experiment.

The effect of choice of energy functional was explored by obtaining results for the $\alpha$ and $\omega$ lattice parameters within the GGA-PBE ~\cite{PBE} and LDA-PZ ~\cite{PZ} exchange-correlation functionals. The results are also presented in table ~\ref{Lattice_parameters}. The best choice of functional for calculating the lattice parameters is GGA, as expected, with only a small difference between the BP and PBE functionals.

Fig. ~\ref{cold_curve} presents the total energy curves for the $\alpha$ and $\omega$ phases calculated with the GGA-BP functional. In these calculations the lattice parameters were allowed to relax independently under the hydrostatic constraint. It can be seen that a third order polynomial is a good approximation for the energy $E(V)$ which allows us to calculate the enthalpies and other thermodynamic quantities that contained derivatives of the energy. Alternative fits to the energy curve do not affect the results significantly. Fig. ~\ref{enthalpy} presents the enthalpy differences between the two phases. From this data we find that $\omega$ is the stable phase at zero temperature and zero pressure in agreement with previous studies ~\cite{Liu} ,~\cite{Alfe}. The phase transition at zero Kelvin (without the zero-point energy contribution) is found to occur at -4.06 $GPa$. This transition shifts to -3.5 GPa if the calculations are performed with the GGA-PBE exchange-correlation functional. Kutepov et al. ~\cite{Kutepov}, Mei et al. ~\cite{Liu} and Hu et al. ~\cite{Alfe} calculated, all with the PBE functional, this transition to occur at -3 GPa, -3.7 GPa and -5 GPa respectively. If we consider a reasonable estimate of the physical uncertainty of the exchange-correlation approximation of ca. 0.5 kJ/mole (5 meV/atom), then the uncertainty in the phase transition pressure increases further and is approximately $-4 \pm 3$ GPa. The volume difference we obtain between $\alpha$ and $\omega$ phases at zero pressure with the BP exchange-correlation functional is 0.246 $\AA^3/atom$. This result agrees well with the previous works of Mei et al. ~\cite{Liu} and Hu et al. ~\cite{Alfe} who predict this volume difference to be 0.23  $\AA^3/atom$ and 0.24  $\AA/^3atom$ respectively, and falls within the range of the experimentally measured data. Zhang et al. ~\cite{Zhang} find the volume change to be 0.33 $\AA/^3atom$ and Tonkov ~\cite{Tonkov} report a value of 0.23 $\AA^3/atom$.

In table ~\ref{lattice_parameter_high_pressure} we present the lattice parameters obtained for the high pressure phases with the GGA-BP functional. It can be seen that the error in the lattice parameters relative to experimentally determined values is less than 1.5\%. We note that in the literature there are no experimental data for the lattice parameters of the $\beta$ phase at the high pressures at which it is stable in our calculations. The positions of the atoms in the unit cell (in crystal coordinates) determined by full relaxation of the lattice parameters and the atomic positions are: (i) $\gamma$: (0.00,0.11,0.25), (0.00,-0.10,0.75), (0.50,0.61,0.25), (0.50,0.40,0.75), (ii) $\delta$: (0.00,0.30,0.25), (0.00,-0.27,0.75), (0.50,0.80,0.25), (0.50,0.23,0.75). In comparison, the experimentally determined positions of the atoms in the unit cell from ~\cite{delta} are: $\gamma$: (0,0.11,0.25), (0,-0.11,0.75), (0.5,0.61,0.25), (0.5,0.39,0.75), $\delta$: (0,0.295,0.25), (0,-0.295,0.75), (0.5,0.795,0.25), (0.5,0.205,0.75). We note that we start our structural relaxation from the experimental values from ~\cite{delta}. As the pressure is increased, the $\delta$ phase evolves and becomes degenerate with the $\beta$ phase with $a/b=c/b=\sqrt{2}$ and atomic positions: (0.00,0.25,0.25), (0.00,-0.25,0.75), (0.50,0.75,0.25), (0.50,0.25,0.75) above ca. 210 GPa, see Fig ~\ref{lattice_par}.

The enthalpy difference of the proposed high pressure phases relative to the $\omega$ phase as a function of pressure are presented in Fig. ~\ref{enthalpy_high_pressure}. The sequence of the stable phases of titanium at high pressure and zero temperature in our calculations is: $\omega \rightarrow \gamma \rightarrow \delta \rightarrow \beta$, as can be seen from Fig. ~\ref{enthalpy_high_pressure}. Again, at high pressure the $\delta$ and $\beta$ phases become degenerate. This result agrees with the experimental results of Akahama et al. ~\cite{delta} except for the stability of the $\beta$ phase at high pressure and does not agree with the results of Ahuja et al. ~\cite{eta}. However, we note that thermal effects not accounted for in the present study may affect this sequence.

\subsection{Elastic constants}
In hexagonal crystals there are 5 independent elastic constant. All the independent elastic constants were calculated using the ElaStic package ~\cite{ElaStic}, in which the second derivative of the energy with respect to the Lagrangian strain is fitted to a polynomial function and from this fit the elastic constants can be obtained. The calculated elastic constants for $\alpha$ and $\omega$ phases are presented in table ~\ref{elastic}. The agreement with the experimental data and other theoretical calculations is of the order of 10\%. The agreement between our calculations and experimental data taken at 4 K ~\cite{Elastic_exp2} is very good, within a few percent for $c_{11}$ $c_{12}$, $c_{33}$, the bulk modulus $B$ and the Poisson ratio $\nu$ (in the Voigt approach). It is also quite good for $c_{13}$, $c_{44}$,(15\% and 18\%, respectively). 

\begin{table}
\caption{Elastic constants for $\alpha$ and $\omega$ phases at zero pressure at zero temperature in GPa: PW- present work, exp1 and exp2 are experimental data from reference ~\cite{Elastic_exp1} and reference ~\cite{Elastic_exp2} (at 4 K) respectively, WIEN2k and QE-US (Quantum-Espresso with ultrasoft pseudopotential) are from reference ~\cite{ElaStic}, VP1 are VASP with PAW from reference ~\cite{Alfe} and VP2 are VASP with PAW from reference~\cite{Hao}.}
\begin{indented}
\item[]\begin{tabular}{@{} l l l l l l l l l }
\br
 &  & $c_{11}$ & $c_{12}$ & $c_{13}$ & $c_{33}$ & $c_{44}$ & $B$ & $\nu$\\
\mr
$\alpha$ & PW & 174.5 & 87 & 78 & 188 & 42 & 114 & 0.325\\
 & exp1 & 160 & 90 & 66 & 181 & 46 & 105 & 0.32\\
 & exp2 & 176 & 87 & 68 & 191 & 51 & 110 & 0.3\\
 & VP1& 175 & 83 & 75 & 196 & 42 & 112 & 0.33\\
 & WIEN2k & 179 & 85 & 74 & 187 & 44 & 112 & 0.31\\
 & QE-US & 190 & 99 & 91 & 213 & 39 & 128 & 0.34\\
$\omega$ & PW & 194 & 85 & 56 & 245 & 53 & 114 & 0.28\\
 & VP1& 199 & 82 & 50 & 247 & 55 & 112 & 0.29\\
 & VP2& 192 & 78 & 55 & 249 & 54 & 112 & \\
\br
\end{tabular}
\label{elastic}
\end{indented}
\end{table}

\subsection{Vibrational properties}

\begin{figure}
\centering
        \includegraphics[scale=0.4]{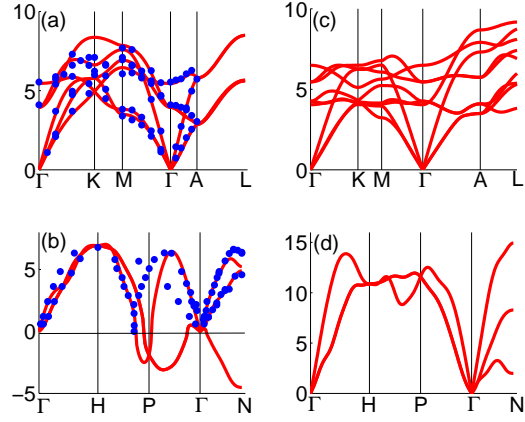}
        \caption{Phonon spectra of titanium in several phases: (a) $\alpha$ at 0 GPa (b) $\beta$ at 0 GPa (c) $\omega$ at 0 GPa (d) $\beta$ at 200 GPa. Dots are experimental data from reference ~\cite{Stassis} for $\alpha$ and reference ~\cite{Petry} for $\beta$}\label{all_spectra}
\end{figure}

\begin{figure}
\centering
        \includegraphics[scale=0.6]{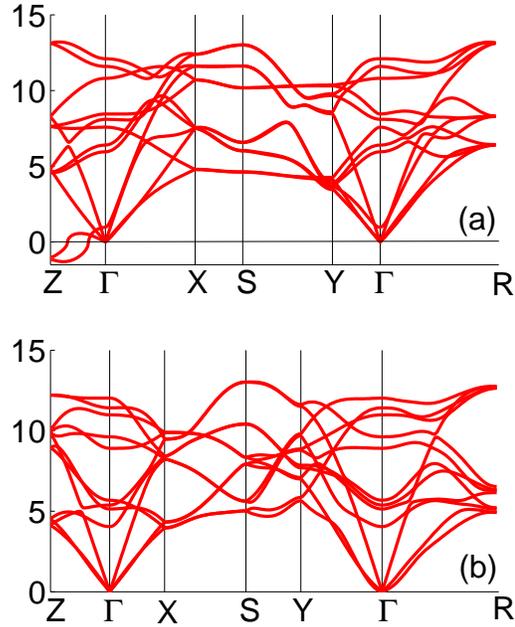}
        \caption{Phonon spectra of titanium in (a) $\gamma$ at 100 GPa (b) $\delta$ at 130 GPa.}\label{all_spectra_high_pressure}
\end{figure}

\begin{figure}
\centering
        \includegraphics[scale=0.5]{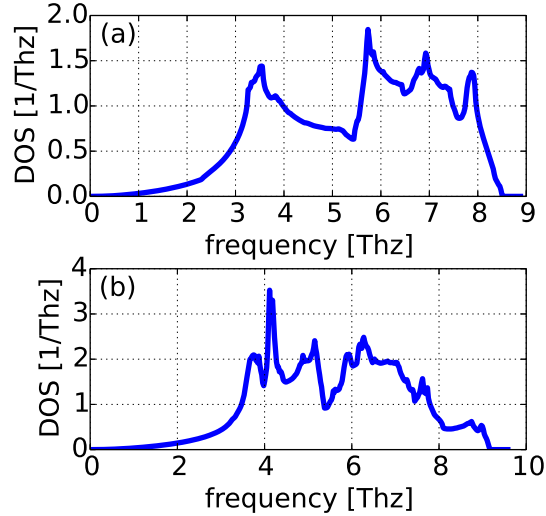}
        \caption{Phonon DOS per unit cell: (a) $\alpha$ (b) $\omega$.}\label{phonon_DOS_subplot}
\end{figure}

The vibrational properties of a crystalline solid are characterized by the phonon spectrum. Figure ~\ref{all_spectra} presents the phonon spectra of the three low pressure phases of titanium at the volumes corresponding to zero pressure and temperature as found in our calculation. The calculated spectrum of the $\alpha$ phase agrees well with the spectrum measured at room temperature ~\cite{alpha_exp_spectrum}. For the $\beta$ phase, the phonon spectrum exhibits imaginary frequencies, which are represented as negative frequencies in the figure, indicating mechanical instability in certain directions. The spectrum in the other directions agrees well with experimental data from neutron scattering at high temperatures ~\cite{beta_exp_spectrum}. However, from this spectrum we cannot calculate thermodynamic properties. The phonon spectra for all phases obtained using GGA-PBE ~\cite{PBE} and LDA-PZ ~\cite{PZ} exchange-correlation approximations did not exhibit any significant differences. 

The phonon spectra of the high pressure phases $\gamma$,$\delta$ and $\beta$ were calculated at representative volumes of 10.85 $\AA ^3/atom$, 9.93 $\AA ^3/atom$ and 8.64 $\AA ^3/atom$ respectively, and are presented in Figs. ~\ref{all_spectra} and ~\ref{all_spectra_high_pressure}. It can be seen that the $\beta$ phase at high pressure becomes mechanically stable and there are no imaginary branches of the phonon spectrum. In addition, at high pressure, we see higher frequencies in the spectrum of the $\beta$ phase than in zero pressure, as expected. The phonon spectrum of the $\gamma$ phase contains imaginary frequencies in the $\Gamma \rightarrow Z (0,0,1/2)$ direction of the reciprocal lattice. These imaginary frequencies represent a mechanical instability of this phase at zero Kelvin. In contrast, we see that the $\delta$ phase does not exhibit imaginary frequencies, indicating its stability at high pressure in agreement with Akahama et al. ~\cite{delta}.

In figure ~\ref{phonon_DOS_subplot} we present the phonon DOS per unit cell of $\alpha$ and $\omega$ phases. In both phases we see a near parabolic behavior of the DOS below a frequency of approximately 3 $Thz$. This parabolic behavior is due to linear acoustic branches well represented by the Debye model. Above 3 $Thz$ we observe a complex behavior of the phonon DOS due to optical branches and non-linearity of the acoustic branches. 
\subsection{Thermodynamic properties}
\begin{figure}
\centering
        \includegraphics[scale=0.45]{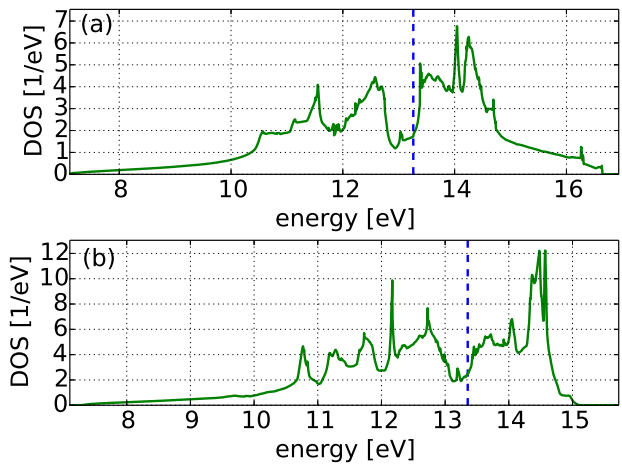}
        \caption{Electronic DOS per unit cell: (a) $\alpha$ (b) $\omega$. The dashed line is the Fermi energy.}\label{subplot}
\end{figure}

\begin{figure}
\centering
        \includegraphics[scale=0.45]{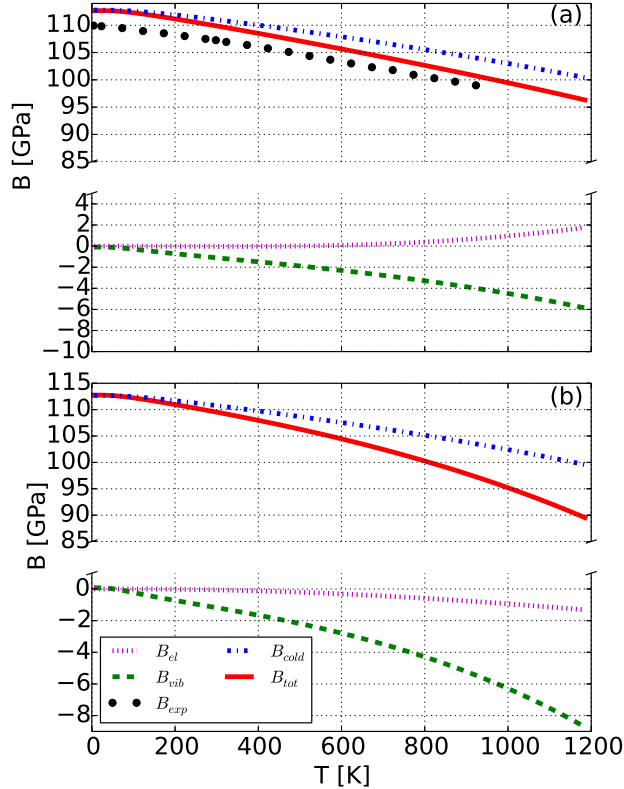}
        \caption{Bulk modulus of (a) $\alpha$ and (b) $\omega$ phases of titanium at constant pressure P=0. Experimental data from reference ~\cite{bulk_modulus}}\label{bulk_subplot}
\end{figure}

\begin{figure}
\centering
        \includegraphics[scale=0.4]{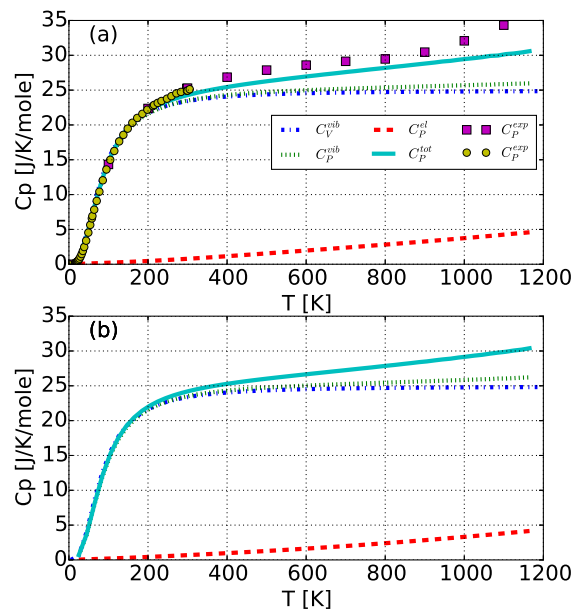}
        \caption{Heat capacity of titanium at constant pressure P=0: (a) $\alpha$ (b) $\omega$. Experimental data for $\alpha$-Ti are in red squares ~\cite{low_T_Cp_exp} and yellow circles ~\cite{NIST}.}\label{Cp}
\end{figure}

\begin{figure}
\centering
        \includegraphics[scale=0.4]{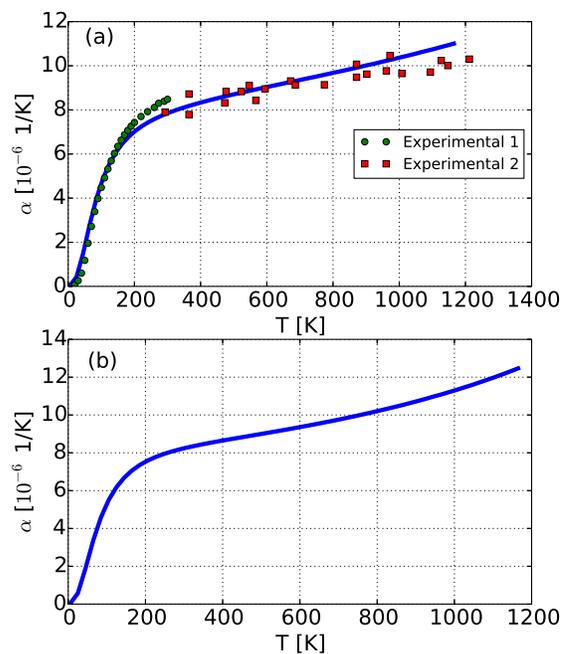}
        \caption{Linear thermal expansion of titanium at constant pressure P=0: (a) $\alpha$ (b) $\omega$. Experimental data is from: 1. reference ~\cite{LTE1} 2. reference ~\cite{LTE2} as cited in ~\cite{Touloukian}.}\label{LTE}
\end{figure}

\begin{figure}
\centering
        \includegraphics[scale=0.4]{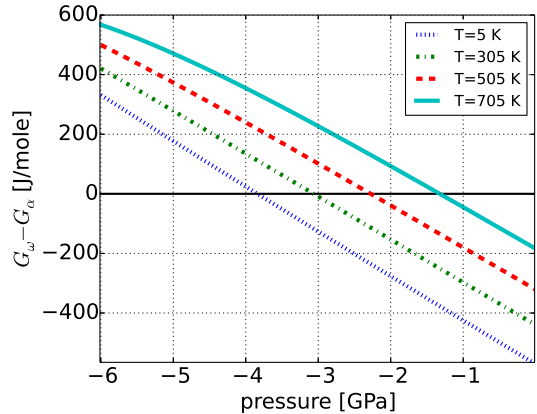}
        \caption{Difference in Gibbs free energies between $\omega$ and $\alpha$ phases. The phase transition pressures are -3.84, -3.06, -2.27 and -1.32 GPa at temperatures 5,305,505 and 705 K respectively.}\label{gibbs}
\end{figure}


\begin{table}
\caption{$\alpha-\omega$ entropy difference (phonons and electrons) at 300 K and 0 pressure (J/K/mole).}
\begin{indented}
\item[]\begin{tabular}{@{} l l l l }
\br
 & GGA-BP & GGA-PBE & LDA-PZ\\
\mr
$\Delta S_{\omega-\alpha}$ & -0.55 & -0.70 & -0.92\\
\br
\end{tabular}
\label{delta_S}
\end{indented}
\end{table}

All thermodynamic properties were calculated from the Helmholtz free energy, equation (~\ref{F}). These results were fitted to a third order polynomial in the volume $V$. From this analytic function the Gibbs free energy and all other thermodynamic properties (at constant pressure as well as at constant volume) can be calculated.

Figure ~\ref{subplot} presents the Kohn-Sham electronic DOS per unit cell for $\alpha$ and $\omega$ phases calculated at volumes that correspond to zero pressure at zero temperature. The DOS of the $\alpha$ phase exhibits a sharp increase slightly above the Fermi level. The DOS in this region increases by a factor of 2 in a very narrow energy range. A similar increase is observable in the $\omega$ phase but it is less sharp. It is even less pronounced per atom since the $\alpha$ phase structure has two atoms per unit cell and the $\omega$ phase has three atoms per unit cell.  

The bulk modulus was calculated from the Helmholtz free energy $F(V,T)$, by the analytical second derivative $(\partial^2 F(V)/\partial V^2)_T$ at each temperature. It is possible to separate the second derivative of the Helmholtz free energy $(\partial^2 F(V)/\partial V^2)_T$ into three contributions: $F_{electron}$, which is the contribution of electronic excitations to the free energy, $F_{phonon}$ which is the contribution of phononic excitations to the free energy and $E$ which is the total energy. We note that all the second derivative components are calculated at the same equilibrium volume $V(T)$.  Figure ~\ref{bulk_subplot} shows the three contributions to the bulk modulus as well as the total bulk modulus, at zero pressure of the two phases as a function of the temperature. In the $\alpha$ phase, the experimentally determined bulk modulus is also shown and it can be seen that the deviation from experiment is about 2\% and is roughly constant with temperature. As expected, the bulk modulus is dominated by the total energy contribution, with the phonon contribution becoming significant above ca. 500 K. 

The heat capacity at constant pressure ($Cp$) for the $\alpha$ and $\omega$ phases is calculated by transforming the Helmholtz free energy to Gibbs free energy. To do so we find the volume $V(P)$ at each temperature by solving $P= -(\partial F(V,T) / \partial V)_T$. Then $Cp=-T\cdot (\partial^2G/\partial T^2)_P$, and the second derivative is calculated numerically at the appropriate pressure. Figure ~\ref{Cp} presents the heat capacity at constant pressure of the $\alpha$ and $\omega$ phases respectively. The calculated heat capacity of the $\alpha$ phase shows good agreement with experimental data at low temperatures. At higher temperatures, the heat capacity shows a deviation from experiment that increases linearly with temperature.  The difference between the calculated heat capacity at constant volume and the calculated heat capacity at constant pressure is small at low temperatures and increases with temperature due to anharmonic effects that are included in the QHA. Overall, the $\alpha$ and $\omega$ phases show very similar electronic heat capacity as well as phonon heat capacity and agree with the results of Nie et al. ~\cite{Nie}. 

The linear volume thermal expansion coefficient, $\alpha = 1/(3V) (\partial V /\partial T)_P$ is calculated from the numerical derivative $(\partial V /\partial T)_P$. It is well known that the thermal expansion of $\alpha$-Ti is anisotropic ~\cite{Souvatzis,Nizhankovskii}. In figure ~\ref{LTE} we present the volume thermal expansion together with experimental data for the $\alpha$ phase with which there is good agreement. The thermal expansion is an anharmonic phenomenon and it seems that the QHA describes well the anharmonic effects in the range of temperature considered, between 0 $K$ to 1200 $K$.

The phonon spectrum may be used to correct the zero Kelvin total energy obtained above to include the effect of zero-point motion of the nuclei. From Figs. ~\ref{gibbs} and ~\ref{enthalpy} we can see the effect of the zero-point energy on the phase transition at low temperatures. The phase transition pressure at 0 $K$ without zero-point energy contribution is -4.06 $GPa$(figure ~\ref{enthalpy}). The phonon contribution shifts the phase transition pressure to -3.84 $GPa$ at 5 $K$ (figure ~\ref{gibbs}), which is also approximately the effect of the zero point energy. This result is in contrast to previous work ~\cite{Liu} where it was suggested that the effect of zero-point energy is to shift the phase transition pressure by 2.2 $GPa$ from -3.7 $GPa$ to -1.5 $GPa$. In addition, we can see the phase transition pressures occur at -3.84, -3.06, -2.27 and -1.32 GPa at temperatures 5, 305, 505 and 705 K respectively in Figs. ~\ref{gibbs}, indicating that within the present calculations the $\alpha$ phase becomes thermodynamically stable only at high temperatures at zero pressure. Similar results are obtained with the PBE exchange-correlation functional.

\begin{figure}
\centering
        \includegraphics[scale=0.4]{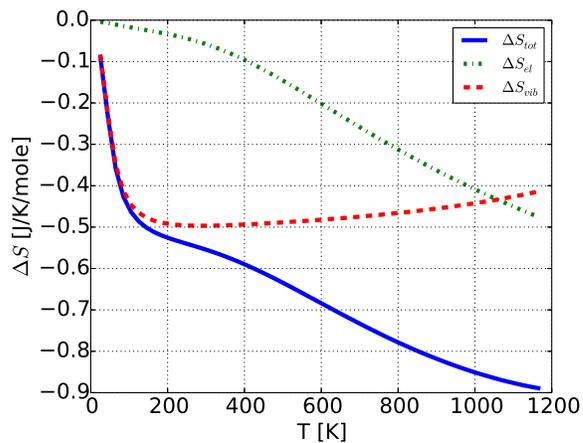}
        \caption{Entropy difference between $\alpha$ and $\omega$ phase as a function of temperature at constant pressure P=0. The total difference (solid line), the phonon contribution (dashed line) and the electronic contribution (dot-dashed line)}\label{delta_s_omega_alpha}
\end{figure}

Figure ~\ref{delta_s_omega_alpha} shows the entropy difference between $\omega$ and $\alpha$ phases. The phononic contribution to the entropy difference increases with temperature up to 200 K and is then approximately constant. In contrast, the electronic contribution to the entropy difference increases monotonically with temperature and becomes dominant at higher temperatures. The choice of exchange-correlation functional affects the value of the entropy difference, even though the effect on the phonon spectrum is small. In table ~\ref{delta_S} we present calculated values of the entropy difference between the $\alpha$ and $\omega$ phases at 300 K and zero pressure for different choices of the exchange-correlation functional. The GGA-BP approximation yields the smallest entropy difference and the LDA-PZ approximation produces the largest.

\section{Discussion.}\label{sec.discussion}
\begin{figure}
\centering
        \includegraphics[scale=0.6]{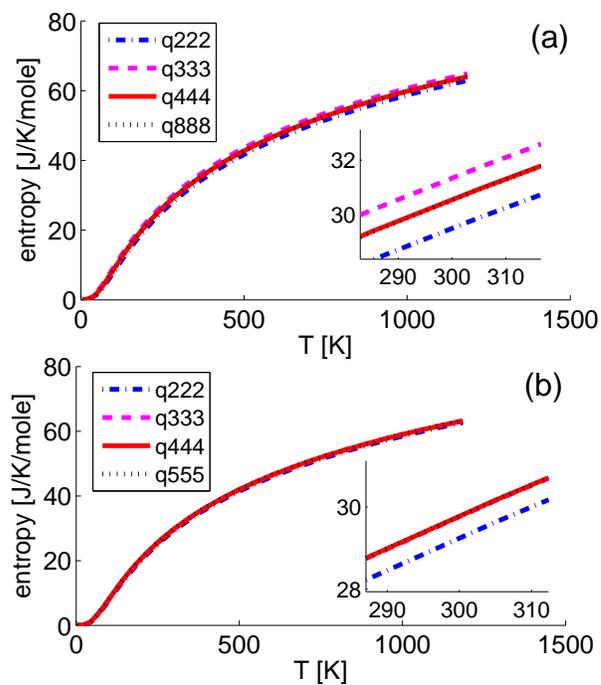}
        \caption{Phonon entropy convergence with respect to the number of q-points: (a) $\alpha$ (b) $\omega$.}\label{entropy}
\end{figure}

\begin{figure}
\centering
        \includegraphics[scale=0.4]{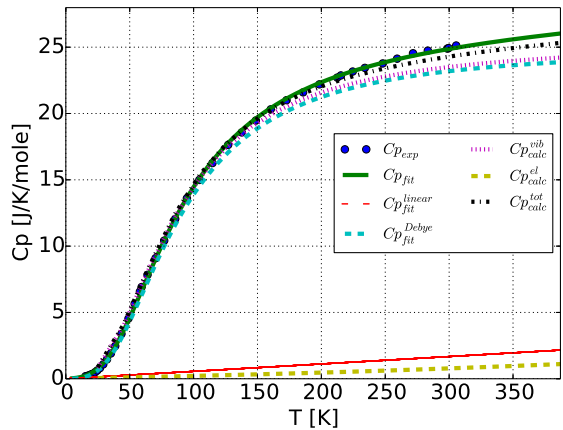}
        \caption{Heat capacity at zero pressure: electronic and phononic calculated heat capacity and experimental results fitted to the Debye heat capacity plus a linear contribution (see text).}\label{Cp_exp}
\end{figure}

We have calculated the fully relaxed lattice parameters of low and high pressure phases of Ti. These are in good agreement with previous theoretical studies and available experimental data. The zero temperature phase stability sequence was found to be $\alpha \rightarrow \omega \rightarrow \gamma \rightarrow \delta \rightarrow \beta$ in agreement with theoretical calculations refs. ~\cite{Hao,Kutepov} and not with reference ~\cite{Liu2}. Our results are in very good agreement with the high pressure experimental study by Akahama et al ~\cite{delta}. We note that the phonon spectrum indicates that the $\gamma$ phase might not be mechanically stable and therefore an additional, undiscovered phase, may exist between $\omega$ and $\delta$. $\delta$ phase becomes structurally and energetically degenerate with the $\beta$ phase at pressures higher than ca. 210 GPa.

The phonon spectra of all six phases were calculated at the relevant pressures. Our results for the $\alpha$, $\omega$ and low pressure $\beta$ phases are in good agreement with previous theoretical and experimental studies. The results reported for the novel high pressure phases $\gamma$, $\delta$ and $\beta$ indicate that the $\beta$ phase is stabilized at the highest pressure calculated. Furthermore, a soft optical mode is obtained in the phonon spectrum of the $\gamma$ phase at 100 GPa in the directions $Y (0,1/2,0) \rightarrow \Gamma \rightarrow R (1/2,1/2,1/2)$. In the directions $\Gamma \rightarrow Z (0,0,1/2)$ both a soft optical mode as well as an acoustic mode have imaginary frequency which indicate a mechanical instability of the $\gamma$ phase. A soft mode is indicative of an imminent change in the crystal structure which is consistent with the small pressure range in which the $\gamma$ phase has been reported to be stable ~\cite{delta}. In the present work this pressure range is 93-112 GPa. However, we find that with increasing pressure the phonon instability of the gamma phase increases and the $Y (0,1/2,0) \rightarrow \Gamma \rightarrow R (1/2,1/2,1/2)$ soft mode is destabilized as well. In the $\delta$ phase at 130 GPa, the soft optical mode in the directions $Y (0,1/2,0) \rightarrow \Gamma \rightarrow R (1/2,1/2,1/2)$ has a similar form to the $\gamma$ phase but the minimum at the $\Gamma$- point is at a higher frequency value, which indicates the mechanical stability of this phase. The enthalpy of the $\delta$ phase is slightly below the enthalpy of the $\beta$ phase.  In addition, we have good agreement with experimental data of the lattice parameters in this phase. From all these results we conclude that the delta phase is stable under hydrostatic pressure in contrast to Mei at al. ~\cite{Liu2} who reported that the delta phase is not stable above 71.8 GPa.

Figure  ~\ref{entropy} exhibits the convergence of the phonon contribution to the entropy as a function of the number of q-points employed in the calculation of the phonon spectrum. It is clear from these figures that the  entropy of the $\omega$ phase converges faster than that of the $\alpha$ phase, probably because of the larger Brillouin zone of the $\alpha$ phase. Therefore, we can conclude that for three q-points, the calculation is converged in the $\omega$ phase to approximately 0.01 $J/K/mol$, which is negligible, but in the $\alpha$ phase only to approximately 0.8 $J/K/mol$. For converged results, it is necessary to work with four q-points in each direction of the reciprocal lattice in the $\alpha$ phase. This is particularly crucial for the calculation of the slope of the phase transition line which is determined by the entropy difference between the two phases, which is of the order of 1 $J/K/mol$.  In the calculations of Mei et al. ~\cite{Liu} and Hu et al. ~\cite{Alfe} the authors did not report on the convergence of their phonon calculations (or resultant quantities such as entropy) with respect to the number of q-points used in the calculations, but they were equivalent to three q-points.

In figure ~\ref{Cp_exp} the low temperature experimental heat capacity ~\cite{Kothen} and the various contributions to the calculated heat capacity are presented. The experimental heat capacity may be fitted to a Debye contribution and a linear contribution: $C_p^{Exp}(T)=C_p^{Debye}(T)+\alpha \cdot T$. In this way we can separate the electronic heat capacity and the phononic heat capacity from the experimental data. Figure ~\ref{Cp_exp} also shows a comparison between the experimental Debye fitted heat capacity $C_p^{Debye}(T)$ to the phononic calculated heat capacity $C_p^{phonons}(T)$ and between the experimental linear fitted heat capacity $\alpha \cdot T$ to the electronic calculated heat capacity $C_p^{electrons}(T)$. At 300 K, $C_p^{phonons}(T)-C_p^{Debye}(T)=0.3$ J/K/mole and $\alpha \cdot T-C_p^{electrons}(T)=0.9$ J/K/mole. These results indicate that most of the error in the heat capacity at high temperatures, as well as in the entropy, originates from the linear contribution to the heat capacity. There are two possible origins of this discrepancy (i) the electronic Kohn-Sham DOS is not a good description of the excited states, especially d-band electrons, (ii) there are other linear contributions to the heat capacity that are not taken into account. We note that this analysis is based on the assumptions that $C_p$ and $C_V$ are approximately the same at these temperatures ,as seen in figure ~\ref{Cp} (a), and that the Debye form for the heat capacity is a good description of the phonon heat capacity. 

\begin{table}
\caption{Debye temperature calculating from $Cv$ ($\theta_{D,Cv}$) and from the elastic constants ($\theta_{D,Elastic}$) at zero pressure. PW refers to present work.}
\begin{indented}
\item[]\begin{tabular}{@{} l l l l l }
\br
 &  & PW & reference ~\cite{Alfe} & reference ~\cite{Elastic_exp2}\\
\mr
$\alpha$ & $\theta_{D,Cv}$ & 350 &  & \\
 & $\theta_{D,Elastic}$ & 403 & 386 & 428\\
$\omega$ & $\theta_{D,Cv}$ & 356 &  & \\
 & $\theta_{D,Elastic}$ & 457 & 439 & \\
\br
\end{tabular}
\label{Debye}
\end{indented}
\end{table}

Table ~\ref{Debye} presents the Debye temperature calculated from both the heat capacity and the elastic constants. It can be seen that there is a significant difference between the Debye temperatures depending on the method of calculation. Employing the Debye temperature calculated from the phononic $C_v$, the entropy difference between $\alpha$ and $\omega$ phases is 0.39 J/K/mole at 300 K. If instead the Debye temperature calculated using the elastic constants is employed the entropy difference between $\alpha$ and $\omega$ phases is 2.83 J/K/mole at 300 K. The entropy difference at 300 K changes by a factor of 7 depending on the method of calculation of the Debye temperature, implying a similar change in the slope $dP/dT=\Delta S/\Delta V$. From these results we conclude that the Debye method is unsuitable to calculate phase transition lines in the present system. 

Finally, we consider the ab-initio construction of the phase transition line between the $\alpha$ and $\omega$ phases. The intersection of this phase transition line with the T=0 K axis was considered in section ~\ref{sec.results}, where we found it to occur at approximately -4 GPa, and other results from density functional theory obtained it in the range of $\pm$1 GPa. If we include the physical uncertainty in the enthalpy difference than based on our estimate the uncertainty is $\pm$4 GPa. The slope of the phase transition line is determined from the Clausius-Clapeyron relation by the ratio of the volume and entropy difference. Although the error in the volume difference is of the order of tens of percent, we find that the value of the entropy difference can vary by a factor of 2 when calculated ab-initio from the phonon calculations and by a larger factor if the Debye approximation is employed. This analysis demonstrates that currently available density functional approximations can provide only rough estimates of transition lines between solid state phases. In particular, they may misidentify the experimentally stable phases in cases were they are nearly degenerate with other low-temperature phases, as is shown here for titanium alpha and omega phases.
\ack{}
The authors acknowledge partial support by the Pazy Foundation.
\section*{References}
\bibliography{refs_no_url_iop}

\end{document}